\documentclass{elsart5p}

\journal{J. Crystal Growth}
\date{21 September 2007}

\usepackage{graphicx}

\usepackage{amssymb}

\begin{document}
\begin{frontmatter}

\title{Phase diagram analysis and crystal growth of \\solid solutions Ca$_{1-x}$Sr$_x$F$_2$}

\author[Berlin]{D. Klimm\corauthref{cor1}}\ead{klimm@ikz-berlin.de},
\author[Berlin]{M. Rabe},
\author[Berlin]{R. Bertram},
\author[Berlin]{R. Uecker}, and
\author[Jena]{L. Parthier}\ead{lutz.parthier@schott.com}

\corauth[cor1]{corresponding author}

\address[Berlin]{Institute for Crystal Growth, Max-Born-Str. 2, 12489 Berlin, Germany}
\address[Jena]{SCHOTT Lithotec AG, Moritz-von-Rohr-Str. 1a, 07748 Jena, Germany}

\begin{abstract}
The binary phase diagram CaF$_2$--SrF$_2$ was investigated by differential thermal analysis (DTA). Both substances exhibit unlimited mutual solubility with an azeotropic point showing a minimum melting temperature of $T_\mathrm{min}=1373^{\,\circ}$C for the composition Ca$_{0.582}$Sr$_{0.418}$F$_2$. Close to this composition, homogeneous single crystals up to 30\,mm diameter without remarkable segregation could be grown by the Czochralski method.
\end{abstract}


\end{frontmatter}


\section{Introduction}
\label{sec:Introduction}

The optical properties of calcium fluoride (CaF$_2$), especially its large transparency range, make this material superior for applications in the short wavelength range, $\lambda<400$\,nm. 
The cubic CaF$_2$ has an isotropic linear refraction index $n$ but shows for very small $\lambda$ nonlinear and nonisotropic intrinsic birefringence.
Intrinsic birefringence scales with $1/ \lambda^2$. At $\lambda=157$\,nm it reaches values up to $11\times10^{-7}$ for CaF$_2$, depending on $\vec{n}$ and polarization, which is one order of magnitude more than acceptable for microelectronic lithography. The intrinsic birefringence is of opposite sign in CaF$_2$ and BaF$_2$ and mixed crystals Ca$_{1-x}$Ba$_x$F$_2$ have been proposed to overcome the problem \cite{Burnett01,Burnett02}. Optical transparency and lattice constants $a_0$ of mixed crystals Ca$_{1-x}$Sr$_x$F$_2$ and Sr$_{1-x}$Ba$_x$F$_2$ were reported by \v{C}ernevskaja \cite{Chernevskaya61,Chernevskaya66}. The dependence $a_0(x)$ was found to be almost linear for both systems.

The binary system CaF$_2$--BaF$_2$ was redetermined recently \cite{Fedorov05} and is quite complicated: The maximal CaF$_2$ solubility in BaF$_2$ is $62\pm5$\,mol\% and the solubility of BaF$_2$ in CaF$_2$ is $8\pm2$\,mol\%. As a result of the miscibility gap, fine lamellar structures with lamella thicknesses $\leq100$\,nm can be found in mixed crystals of intermediate composition and the material must be regarded as non useful for lithography applications.

Some of us reported recently on first promising results with Ca$_{1-x}$Sr$_x$F$_2$ mixed crystals \cite{Parthier05}. The difference of cationic radii in this system is smaller than in Ca$_{1-x}$Ba$_x$F$_2$ (Ca$^{2+}$: 126\,nm, Sr$^{2+}$: 140\,nm, Ba$^{2+}$: 156\,nm) and larger mutual solubility of the fluorides crystallizing in the same (fluorite) structure can be expected. Indeed it could be found in the present work that CaF$_2$ and SrF$_2$ exhibit unlimited mutual solubility without miscibility gap.


\section{Experimental}
\label{sec:Experimental}

A NETZSCH STA 409C thermal analyzer with graphite heater and standard DTA sample holder (thermocouples Pt90Rh10--Pt) was used for DTA measurements with heating/cooling rates of $\pm10$\,K/min in graphite crucibles without lid in an 80\,ml/min argon flow of 99.999\% purity. Pieces of CaF$_2$ single crystal (UV optical quality), of crystalline SrF$_2$ (Merck, 99.99\%), or mixtures of them, respectively, with a total mass of $80-120$\,mg were used as samples. In total 10 compositions Ca$_{1-x}$Sr$_{x}$F$_{2}$ $(0\leq x\leq1)$ were investigated. Mixing was obtained by a first heating run up to $1510^{\,\circ}$C and subsequent cooling to $800^{\,\circ}$C. All DTA curves that are discussed here were obtained in a following second DTA heating run and raw data for the temperature $T$ were stored in the DTA data files.

Crystal growth was performed in a commercial Czochralski puller (Cyberstar, Grenoble, France) with \emph{rf} heating and automatic diameter control. The experiments were performed with Ca$_{1-x}$Sr$_x$F$_2$ batches of $140-150$\,g from graphite crucibles with 120\,mm diameter in a gas mixture of 95\% Ar (99.999\% purity) with 5\% CF$_4$ (99.995\% purity).

The chemical composition of Ca$_{1-x}$Sr$_x$F$_2$ mixed crystals and of melt remaining in the growth crucibles was measured with an ICP-OES (Inductively Coupled Plasma--Optical Emission Spectroscope) ``IRIS Intrepid HR Duo'' (Thermo Elemental, USA). The precision is $\approx3$\% relative standard deviation (R.S.D.) for concentrations above background equivalent concentration (BEC).


\section{Results}
\label{sec:Results}
\subsection{Thermal analysis and thermodynamic assessment}
\label{sec:DTA}

\begin{figure}[ptbh]
\begin{center}
\includegraphics[width=0.4\textwidth]{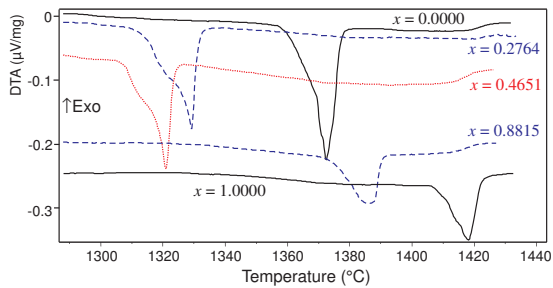}
\caption{DTA melting peaks of Ca$_{1-x}$Sr$_{x}$F$_{2}$ with (from top to bottom) $x=0.0000$ (pure CaF$_2$), $x=0.2764$, $x=0.4651$, $x=0.8815$, and $x=1.0000$ (pure SrF$_2$). (Raw data, not yet calibrated)}
\label{fig:DTA-curves}
\end{center}
\end{figure}

The 2$^\mathrm{nd}$ DTA heating curves around the melt peaks of samples Ca$_{1-x}$Sr$_x$F$_2$ with 5 (of 11 in total) different molar compositions $x$ are shown in Fig.~\ref{fig:DTA-curves}. It should be remarked, that these curves are not yet corrected for $T$. The calibration of $T$ was performed after the measurements with the melting points of both pure substances $T_\mathrm{f}(x=0)=1418^{\,\circ}$C and $T_\mathrm{f}(x=1)=1477^{\,\circ}$C \cite{ChemSage4_20}. $T_{\textrm{f}}$ drops from both pure end members (solid lines) to lower values. The minimum melting temperature (taken from the extrapolated onset of the peak) was observed for $x=0.4651$ (dotted curve in Fig.~\ref{fig:DTA-curves}), but the peak onset was only slightly higher (by $\approx1$\,K) for the neighboring compositions $x=0.4120$ or $x=0.5185$, respectively, that are not shown in Fig.~\ref{fig:DTA-curves}. Obviously, $T_{\textrm{f}}(x)$ has a local minimum (azeotropic point) for some composition near $x=0.4651$.

After measurement the contents of the DTA crucible was always molten to a clear sphere. Inspection with an optical microscope did not reveal obvious inclusions.

\begin{figure}[ptbh]
\begin{center}
\includegraphics[width=0.4\textwidth]{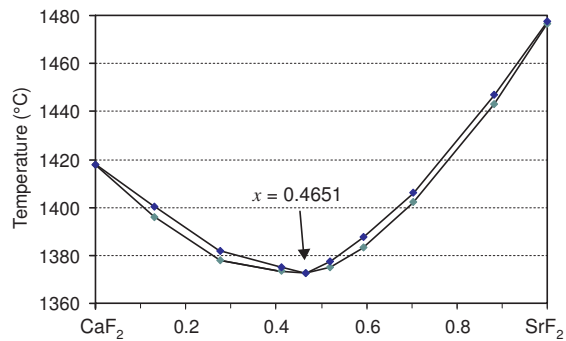}
\caption{Binary system CaF$_{2}$ -- SrF$_{2}$ as obtained from DTA curves (Fig.~\ref{fig:DTA-curves}) after $T$ correction.}
\label{fig:PD-meas}
\end{center}
\end{figure}

The experimental points determining the solidus line in the preliminary phase diagram presented on Fig.~\ref{fig:PD-meas} were taken from the extrapolated onsets ($T_{\mathrm{onset}}$) of the 10 measured compositions after $T$ correction at the $T_{\textrm{f}}$ of both end members (as written above $T_{\mathrm{sol}}=T_{\mathrm{onset}}$). Unfortunately, the determination of the liquidus from DTA peaks is not as straightforward as the determination of the solidus. It turned out that the width of the DTA peaks

\begin{equation}
\mathit\Delta T(x) = T_{\mathrm{offset}}(x) - T_{\mathrm{onset}}(x)
\label{eq:width}
\end{equation}

was 8.1\,K for pure CaF$_2$ ($x=0.0000$) as well as for $x=0.4651$, respectively. For pure SrF$_2$ ($x=1.0000$) a slightly larger $\mathit\Delta T = 9.3$\,K was found. The experimental points determining the liquidus in Fig.~\ref{fig:PD-meas} were calculated by

\begin{eqnarray}
\nonumber T_\mathrm{liq}(x) &=& T_\mathrm{onset}(x)+\mathit\Delta T(x)-8.1\mathrm{\,K} \\
                            &=& T_{\mathrm{offset}}(x)-8.1\mathrm{\,K} \label{eq:liquidus}
\end{eqnarray}

with a maximum measured $\mathit\Delta T=12.3$\,K that was observed for $x'=0.5932$ and for $x''=0.1317$. Procedure (\ref{eq:liquidus}) results in a difference of 1.2\,K between the solidus and liquidus for pure SrF$_2$ which is regarded as the experimental error.

The thermodynamic assessment was started with the data for the pure end members, CaF$_2$ and SrF$_2$, from the ChemSage database SPS96TO2 \cite{ChemSage4_20}. Three mixed phases were calculated, each with species CaF$_2$ and SrF$_2$ and for the whole composition range $0\leq x\leq1$:

\begin{description}
\item[gas:] IDMX model (ideal mixing)
\item[liquid:] SUBI (two-sublattice ionic solution)
\item[(Ca,Sr)F$_2\langle$ss$\rangle$:] SUBL (ideal sublattice solution)
\end{description}

Both non-ideal models of mixing take into account that anions and cations are mixing independently. The Gibbs free energy of both non-ideal mixed phases ($G_{\mathrm{mixed}}$) with interacting Ca$^{2+}$ and Sr$^{2+}$ occupying the cation lattice in presence of F$^-$ on the anion lattice can then be described as the sum of the summands $G_0$ (weighed contribution of the pure components), $G_{\mathrm{id}}$ (contribution of an ideal mixture), and $G_{\mathrm{ex}}$ (contribution of non-ideal interaction):

\begin{eqnarray}
G_{\mathrm{mixed}} &=& G_{0} + G_{\mathrm{id}} + G_{\mathrm{ex}} \label{eq:G1} \\
G_{0}           &=& (1-x)G^{\mathrm{Ca}} + x G^{\mathrm{Sr}} \\
G_{\mathrm{id}} &=& RT\left((1-x)\ln(1-x) + x \ln x\right)\\
G_{\mathrm{ex}} &=& (1-x)\: x \left(a_{1}+b_{1}T+d_{1}T^{2}+a_{3}(2 x - 1)\right) \label{eq:G4}
\end{eqnarray}

(All data are given here in J/mol or in K.) $G^\mathrm{Ca}$ or $G^\mathrm{Sr}$, respectively, are the database standard values for the pure fluorides \cite{ChemSage4_20}. Four parameters for the solid phase and two parameters for the liquid phase were found sufficient for a satisfactory fit of the experimental data using the ChemSage parameter optimization module. These data are $a_{1}=2046.1957$, $b_{1}=-2.3570417$, $d_{1}=-0.0045433252$, $a_{3}=-844.93984$ for the mixed crystal (Ca,Sr)F$_2\langle$ss$\rangle$ and $a_{1}=-43.004897$ and $b_{1}=-11.559429$ for the liquid phase. The deviation from ideality is much smaller for the liquid as compared to the solid.

\begin{figure}[htbp]
\begin{center}
\includegraphics[width=0.45\textwidth]{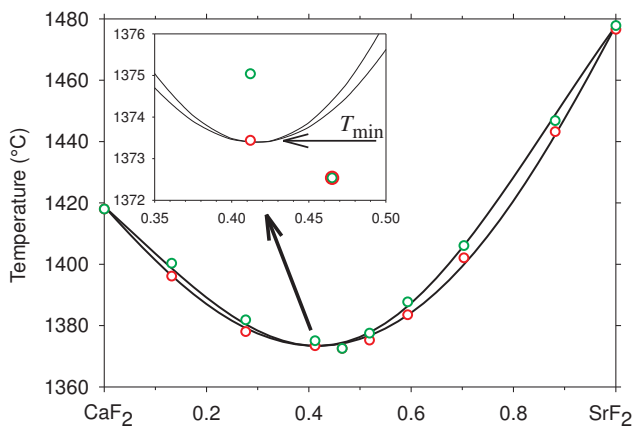}
\caption{Thermodynamic assessment of the CaF$_2$--SrF$_2$ phase diagram (lines) together with the experimental points for liquidus and solidus from Fig.~\ref{fig:PD-meas}.}
\label{fig:PD-calc-meas}
\end{center}
\end{figure}

Fig.~\ref{fig:PD-calc-meas} compares the result of the assessment (\ref{eq:G1})--(\ref{eq:G4}) with the experimental points. The remaining difference is smaller 1\,K --- indicating that the thermodynamic models for the $G_{\mathrm{ex}}$ are reasonable. Note that the azeotropic point of the assessment $x(T_\mathrm{min})\approx0.418$ does not coincide with the lowest experimental point that was found at $x=0.4651$ (insert in Fig.~\ref{fig:PD-calc-meas}).


\subsection{Crystal Growth and Characterization}
\label{sec:Crystal}

\begin{figure}[htbp]
\begin{center}
\includegraphics[width=0.40\textwidth]{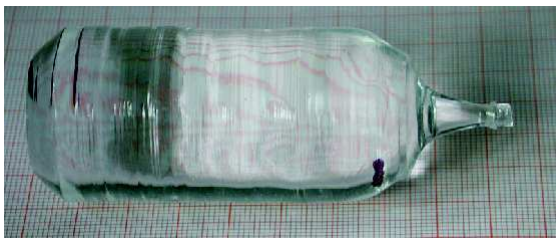}
\caption{CaF$_2$--SrF$_2$ mixed crystal grown at the composition Ca$_{0.585}$Sr$_{0.415}$F$_2$ with 30\,mm diameter and 50\,mm length of the cylinder.}
\label{fig:Xtal}
\end{center}
\end{figure}

For the crystal growth experiments powdered CaF$_2$ and SrF$_2$ from GFI ``crystal grade" was used. The SrF$_2$ concentration was $0.38\leq x\leq0.479$ and the pulling rate $1.5-2$\,mm/h for crystals with 18\,mm diameter. Some crystals were grown with larger diameter up to 30\,mm (Fig.~\ref{fig:Xtal}).

\begin{figure}[htbp]
\begin{center}
\includegraphics[width=0.35\textwidth]{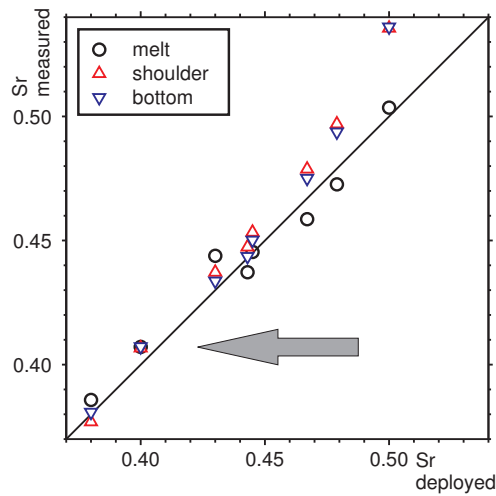}
\end{center}
\caption{Strontium concentration (expressed as molar fraction $x$ of SrF$_2$) in the starting composition ($x_0$) compared to measured concentrations in the shoulder and bottom of Czochralski crystals and the remainder of the melt ($x_\mathrm{meas}$). The straight line is $x_\mathrm{meas}=x_0$.}
\label{fig:ICP-OES}
\end{figure}

The concentrations of calcium ($1-x$) and strontium ($x$) were measured by ICP-OES at the shoulder and bottom, respectively, of several 18\,mm crystals and in the remaining melt. It must be expected that only for growth runs without segregation $x$ is constant for all 3 measurements, and moreover that $x$ equals the Sr concentration in the starting material. Fig.~\ref{fig:ICP-OES} compares the Sr concentration in the starting material (``Sr deployed" = $x_0$) with the measured $x_\mathrm{meas}$. It can be seen that only for $x=0.40$ the Sr concentration is the same for the 2 positions in the crystal and for the melt. The small positive deviation from the ideality line $x_\mathrm{meas}=x_0$ is attributed to experimental errors of the concentration measurement.


\section{Summary}
\label{sec:Summary}

It could be confirmed by DTA that CaF$_2$ and SrF$_2$ show complete miscibility in both liquid and solid phases. Mixed crystals Ca$_{1-x}$Sr$_x$F$_2$ were grown in the concentration range $0.38\leq x\leq0.479$ by the Czochralski method. Both DTA and crystal growth experiments showed that the liquidus and solidus lines of the binary phase diagrams meet at a common minimum (azeotropic point) $x\gtrsim0.4$ with $T_\mathrm{min}=1373^{\,\circ}$C. A Thermodynamic assessment of the DTA melting curves showed the azeotropic point at $x(T_\mathrm{min})\approx0.418$. Homogeneous Ca$_{0.582}$Sr$_{0.418}$F$_2$ single crystals can be grown at this point from the melt as no segregation must be expected.


\end{document}